\documentclass[a4paper]{article}
\usepackage{ASVspoof2021}
\usepackage{multirow}
\usepackage{array}
\usepackage{xcolor}
\usepackage{pifont}
\usepackage{booktabs}
\newcolumntype{P}[1]{>{\centering\arraybackslash}p{#1}}
\usepackage{epsfig,amssymb,amsmath}

\usepackage{url}
\usepackage{graphicx}
\usepackage{amsmath}
\usepackage{hyperref}
\usepackage{booktabs}
\usepackage{bm}
\usepackage{amssymb}
\usepackage{color}
\usepackage{tablefootnote}

\usepackage{xcolor}

\ninept

\title{UR Channel-Robust Synthetic Speech Detection System for ASVspoof 2021}

\name{Xinhui Chen\textsuperscript{*}, You Zhang\textsuperscript{*}\thanks{* The first three authors have equal contributions, presented in alphabetical order.}, Ge Zhu\textsuperscript{*}, Zhiyao Duan}

\address{Audio Information Research Lab, University of Rochester, Rochester, NY, USA \\
{\small \tt you.zhang@rochester.edu} }
\begin{document}
\maketitle

\begin{abstract}
In this paper, we present UR-AIR system submission to the logical access (LA) and the speech deepfake (DF) tracks of the ASVspoof 2021 Challenge. The LA and DF tasks focus on synthetic speech detection (SSD), i.e. detecting text-to-speech and voice conversion as spoofing attacks. Different from previous ASVspoof challenges, the LA task this year presents codec and transmission channel variability, while the new task DF presents general audio compression. Built upon our previous research work on improving the robustness of the SSD systems to channel effects, we propose a channel-robust synthetic speech detection system for the challenge. To mitigate the channel variability issue, we use an acoustic simulator to apply transmission codec, compression codec, and convolutional impulse responses to augmenting the original datasets. For the neural network backbone, we adopt Emphasized Channel Attention, Propagation and Aggregation Time Delay Neural Networks (ECAPA-TDNN) as our primary model. We also incorporate one-class learning with channel-robust training strategies to further learn a channel-invariant speech representation. Our submission achieved EER 20.33\% in the DF task; EER 5.46\% and min-tDCF 0.3094 in the LA task.

\end{abstract}

\noindent\textbf{Index Terms}: synthetic speech detection,
anti-spoofing,
deepfake,
data augmentation,
channel robustness,
ASVspoof 2021
\section{Introduction}
\label{sec:intro}
Automatic speaker verification (ASV) systems have been widely used in biometric authentication, where a person's identity is automatically verified with acceptance or rejection by analyzing speech utterances~\cite{markowitz2000voice, hansen2015speaker}. However, past research has shown that the ASV system is vulnerable to spoofing attacks, where attackers pretend to  be  the target speaker by presenting false but similar-to-bona-fide speech trials~\cite{wu2015spoofing}. Anti-spoofing has been a research topic to deal with this issue.

Synthetic speech detection (SSD) aims to distinguish the text-to-speech (TTS) and voice conversion (VC) attacks from bona fide speech. It is an important issue in anti-spoofing since the speech synthesis techniques are fast developing and will cause more threats to speaker verification systems~\cite{7400997, kamble2020advances}. In a recent study~\cite{muller2021human}, some synthetic speech is even perceptually non-distinguishable from bona fide speech.


One issue in SSD is the generalization ability towards unseen spoofing attacks during training. Efforts have been put into the study of generalization with a focus on feature engineering~\cite{sahidullah2015comparison}, backend model~\cite{monteiro2020generalized}, loss function~\cite{zhang2021one, wang2021comparative}, data augmentation~\cite{chen2020generalization}, and model ensemble~\cite{monteiro2020ensemble}.
According to the result of the ASVspoof 2019 Challenge, for the LA scenario, very few systems deliver EERs below 5\%~\cite{todisco2019asvspoof}, which suggests the necessity of better countermeasures for synthetic speech attacks. 
Besides, there is a big performance gap between training and evaluation, suggesting generalization issues to unseen attacks. 

Another issue is the robustness to the variability of recording conditions, including background noise and channel effects. The performance degradation of state-of-the-art spoofing detectors under unseen conditions has been revealed in some cross-dataset studies for SSD~\cite{das2020assessing, das2020predictions, hanilci2016spoofing, zhang2021empirical}. Of all the factors presenting cross-dataset effects, we tested and verified the channel effect as a reason for performance degradation of SSD systems in~\cite{zhang2021empirical}. Without considering these effects, the systems may fail to generate robust results under different variations. 


 

The logical access (LA) track of the ASVspoof 2021 Challenge~\cite{2021asvspoof} also aims at designing generalized countermeasures that are robust to codec and transmission channel variability. 
In addition, a new SSD task called speech deepfake (DF) is proposed in ASVspoof 2021 to detect fake audio without ASV systems. For this task, developing systems that can be applied in multiple and unknown conditions such as audio compression is also critical. Furthermore, both LA and DF use attacks unseen during training for evaluation, as in ASVspoof 2019.
This makes the ASVspoof 2021 challenge a great test bench for evaluating the generalization ability and channel robustness of SSD systems.

In this paper, we develop a system based on our previous work for addressing the generalization ability with one-class learning~\cite{zhang2021one} and the robustness issue with channel-robust strategies~\cite{zhang2021empirical}.
Our main contributions are threefold:
\begin{itemize}
\item We design several data augmentation methods depending on the task focus: for the DF task, we applied Fraunhofer MPEG layer III (MP3) and advanced audio codec (AAC) audio compression codecs; for the LA task, we introduce three common transmission codec degradation including landline, cellular, and Voice over Internet Protocol (VoIP). We also propose to involve device impulse responses to the above two augmentation pipelines as an additional step to remove the dataset bias introduced by ASVspoof 2019 training dataset.
\item We employ a state-of-the-art backbone in several speaker recognition tasks, ECAPA-TDNN \cite{desplanques2020ecapa} as the primary architecture, and we ensemble its variants to further improve the final performance. To the best of our knowledge, we are the first to have ECAPA-TDNN in the ASV anti-spoofing task.
\item We adapt the channel-robust training strategies proposed in~\cite{zhang2021empirical} to pair with the new data augmentation methods.
\end{itemize}


\section{System Description}
\label{sec:sys}

\subsection{Data Augmentation}
\label{ssec:dataaug}

In our previous work~\cite{zhang2021empirical}, we verified that channel effects are an important reason for cross-dataset performance degradation, and data augmentation methods are shown effective to improve the cross-dataset performance of anti-spoofing systems. To address the compression variability in the DF task and the transmission variability in the LA task, we design different augmentation methods correspondingly.

\begin{figure}[]
  \centering
  \includegraphics[width=\linewidth]{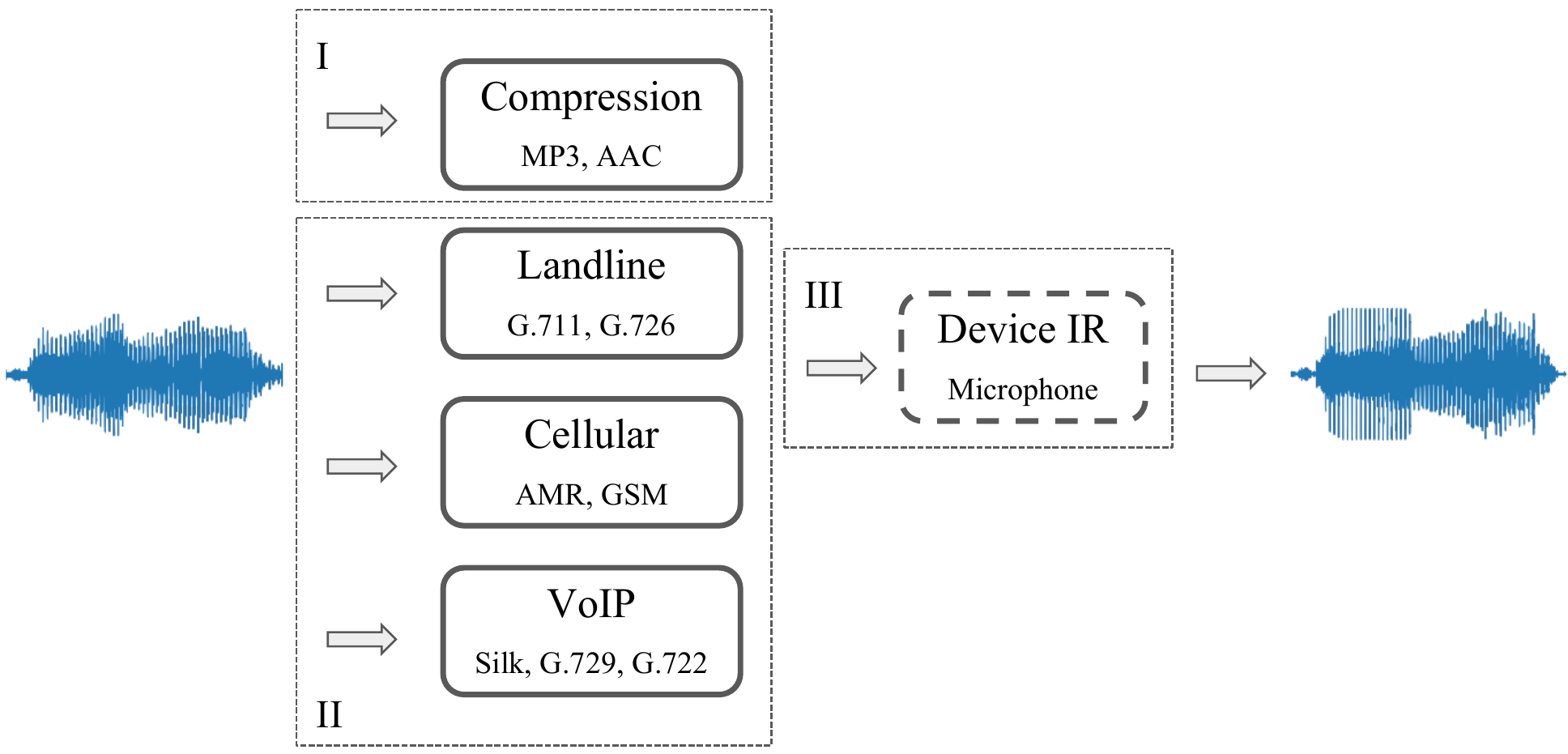}
  \caption{Illustration of our proposed data augmentation pipeline. The first step is codec augmentation, where part I is for the DF task and part II is for the LA task. The second step is device augmentation, which is optional for both tasks.
}
  \label{fig:aug}
\end{figure}

Figure~\ref{fig:aug} shows the pipeline for our proposed data augmentation methods. We utilize four groups of different lossy codec and several impulse responses collected in the open-source acoustic simulator~\cite{ferras2016large}. Different from the previous work where we only explored the device IRs, we first apply codec degradation, i.e., compression and transmission, and then optionally convolve them with recording device IRs. 

\subsubsection{Codec Augmentation}
\label{sssec:codec}

\textbf{Compression Augmentation for DF.}
Compression refers to compression-decompression algorithms that are used to process the audio files, the deepfake speech may be spread across multiple conditions such as general audio compression rather than telephony. As shown in Figure~\ref{fig:aug} part I, for our compression augmentation, we apply MP3 and AAC codecs in the acoustic simulator.

\textbf{Transmission Augmentation for LA.}
Transmission refers to codecs that the speech data is transmitted across before being processed by anti-spoofing systems. The transmission process would introduce degradation due to different bandwidths packet loss and sampling frequencies. However, the degradation should not be considered as artifacts that we aim to detect for discriminating spoofing attacks.
As shown in Figure~\ref{fig:aug} part II, for transmission degradation, we apply landline, cellular and VoIP codec groups, which are the most commonly used transmission channels. To be specific, the landline codec includes $\mu$-law and A-law companding and adaptive-differential PCM. Cellular includes GSM and narrow-band and wide-band advance multi-rate (AMR-NB and AMR-WB). And VoIP includes ITU G.722, ITU G.729 standards besides SILK and SILK-WB. All of the above-mentioned codecs are equipped with different bitrate options. After going through these different transmission codecs, we resample all of the resulting audio files with 16 kHz for further processing. 

\subsubsection{Device Augmentation}
\label{sssec:device}
As shown in our previous study~\cite{zhang2021empirical} where we only applied device augmentation, it is beneficial to introduce various device impulse responses to mitigate the device bias induced by the ASVspoof2019LA training dataset and work robustly across channel variations. As shown in Figure~\ref{fig:aug} part III, 
we propose to convolve the original audio data with recording device IRs after applying compression or transmission augmentation.

\begin{figure}[]
  \centering
  \includegraphics[width=0.76\linewidth]{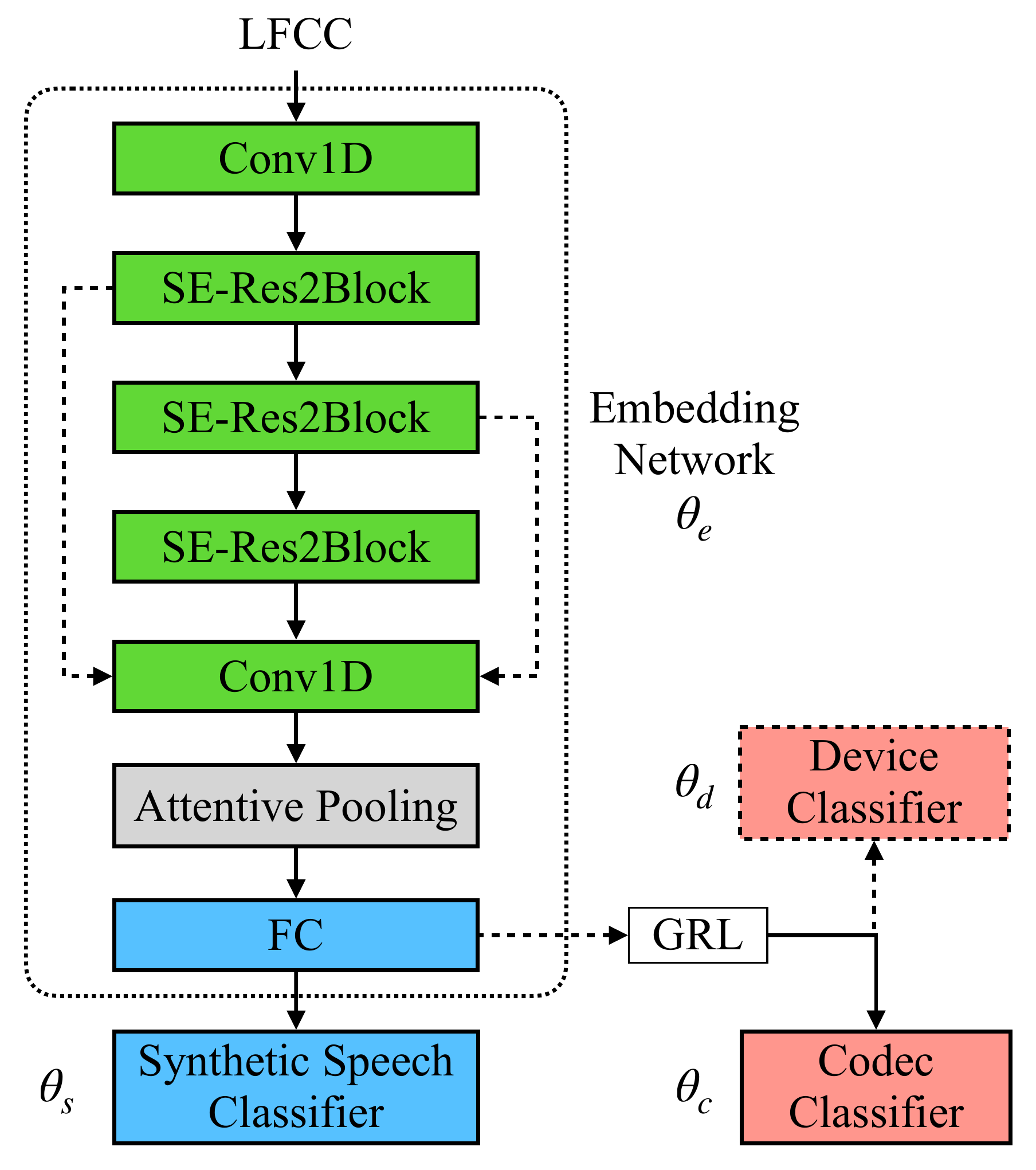}
  \caption{Proposed model architecture with a channel-robust training strategy. Dashed arrows represent optional operations.
}
  \label{fig:arch}
\end{figure}

\subsection{Model Architecture: ECAPA-TDNN}
\label{ssec:model}

Time-delay neural networks (TDNN) \cite{waibel1989phoneme, peddinti2015time} were originally proposed in modeling long-range temporal dependencies and have long been successfully used in speech and speaker recognition tasks \cite{han2021multistream, snyder2018x}. It is similar to the 1-dimensional dilated convolutional neural network (CNN) where only some specific frames in the convolution context window are used. Compared to ordinary CNN, it requires fewer parameters and achieves better results when processing the same width of temporal context~\cite{peddinti2015time, zeinali2019improve}. 

In this paper, we employ an improved version of TDNN called ECAPA-TDNN~\cite{desplanques2020ecapa}, which has introduced several enhancements to improve the original TDNN including Squeeze-Excitation (SE) blocks, multi-scale Res2Net features, multi-layer feature aggregation, and channel-dependent attentive statistics pooling. The model architecture is illustrated in Figure~\ref{fig:arch}. 
Specifically, the 1-dimensional SE-Res2Net layer first utilizes SE blocks to re-scale the channels of feature maps with global context information in the convolutional blocks. The Res2Net architecture further improves the performance and simultaneously reduces the total number of parameters through grouped convolutions, this layer is similar to a recent successful anti-spoofing system~\cite{li2021replay}. Another key component in ECAPA-TDNN is the aggregation of multi-level features through skip connections, as illustrated with the dashed lines between the green blocks in Figure~\ref{fig:arch}.

With the combination of the above-mentioned modules, the ECAPA-TDNN architecture recently achieved the best performance in all three tracks of speaker verification in VOXSRC~\cite{nagrani2020voxsrc} and the text-independent task of SdSV Challenge on short-duration speaker verification~\cite{zeinali2020sdsv}. As the SSD task also aims to learn a discriminative speaker embedding to distinguish spoofing attacks and considers temporal contextual modeling, we believe that the ECAPA-TDNN architecture is also a good choice for the deep learning backbone.

In this paper, we studied and compared six setups of ECAPA-TDNN, shown in Table~\ref{tab:ecapav}, by varying the number of filters of convolutional layers, and whether or not to insert the multi-layer feature aggregation (MFA) block and channel- and context-dependent statistics pooling (CCSP). Although the channel- and context-dependent vector before pooling only marginally improves speaker recognition performance in the original ECAPA-TDNN system \cite{desplanques2020ecapa}, we include this operation as an option here as it could be beneficial in our application.

\begin{table}[ht!]
\setlength{\tabcolsep}{4.5pt}
\renewcommand{\arraystretch}{1.1}
  \centering
  \caption{ECAPA-TDNN backbone variants}
  \begin{tabular}{P{40pt}|P{50pt} | P{40pt}| P{40pt}}
    \hline\hline
    \textbf{System} & \textbf{\# of Filters}   & \textbf{CCSP}& \textbf{MFA}\\
    \hline
    ECAPA-1& \multirow{3}{*}{512} & \ding{51}  & \ding{55}\\

    ECAPA-2&   & \ding{55}  & \ding{51}\\

    ECAPA-3 &   & \ding{51}  & \ding{51}\\
    \hline
    ECAPA-4& \multirow{3}{*}{1024} & \ding{51}  & \ding{55}\\

    ECAPA-5&  & \ding{55}  & \ding{51}\\

    ECAPA-6&  & \ding{51}  & \ding{51}\\
    \hline\hline
  \end{tabular}
  \label{tab:ecapav}
\end{table}

\subsection{Training Strategy}
\label{ssec:strategy}
To improve the generalization ability, the idea of one-class learning~\cite{zhang2021one} with OC-Softmax as the loss function is adopted, where two different margins are introduced for compacting the bona fide speech in the embedding space and isolating the spoofing attacks, to prevent overfitting known conditions. 

To improve the channel robustness, we adopt the channel-robust training strategies proposed in~\cite{zhang2021empirical} to take advantage of the augmented data. 
As the new evaluation dataset in ASVspoof 2021 introduces new channel variability, we assume that it has different distributions from the training data.

\begin{figure}[!htbp]
  \centering
  \includegraphics[width=\linewidth]{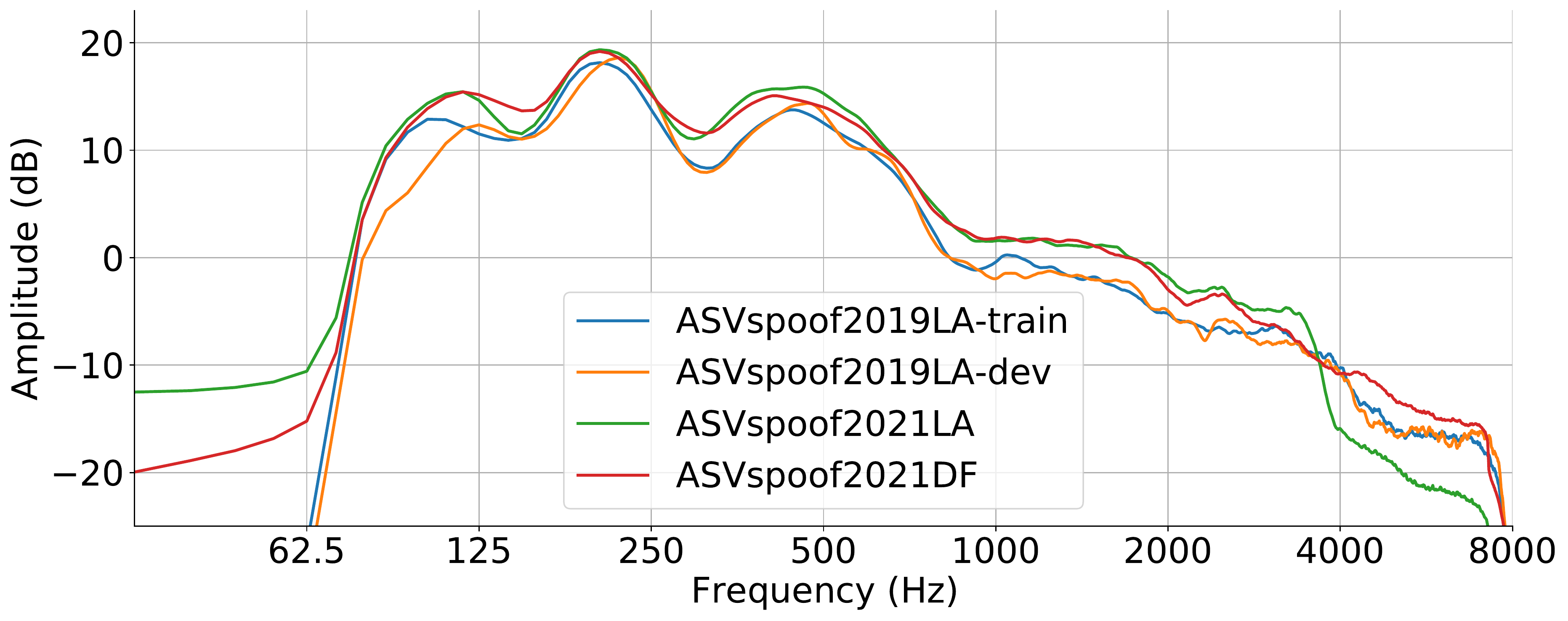}
  \caption{Comparison of average magnitude spectra of utterances across different sets of data. 
}
  \label{fig:aver_spec}
\end{figure}

In Figure~\ref{fig:aver_spec}, we compare the average magnitude spectrum of the LA and DF datasets of the ASVspoof 2021 challenge with the ASVspoof 2019 training data across all utterances of each dataset to illustrate the channel effects. We can see that the average spectra are very different for LA, especially below 62.5 Hz and above 4 kHz. The low-frequency component could be caused by some channel noise. And the high-frequency component could be missing for some utterances, due to the fact that some transmission codecs, such as cellular or landline sampled at 8kHz, will lose the spectral information above Nyquist frequency. For DF, the main difference exists in frequencies below 62.5 Hz, which could also be caused by channel noise. With this observation, we verify that channel mismatch does exist across the training and evaluation data for this challenge. As such, with this observation, we assume that the channel-robust strategies would improve the performance.

We follow the augmentation structure in~\cite{zhang2021empirical} and use plain augmentation (AUG) training and adversarial (ADV-AUG) training strategies. As illustrated in Figure~\ref{fig:arch}, in the AUG strategy, only the embedding network, parameterized by $\theta_{e}$, and the synthetic speech classifier parameterized by $\theta_{s}$ are utilized. In ADV-AUG strategy, to fit the data augmentation methods in Section~\ref{ssec:dataaug}, we design two additional different classifiers, parameterized by $\theta_{c}$ and $\theta_{d}$ respectively, to deal with two categories of augmentation, codec and device. 
 The overall training objective is:
\begin{equation*}
\begin{aligned}
(\hat{\theta}_{e}, \hat{\theta}_\textit{s}) &=\underset{\theta_{e}, \theta_\textit{s}}{\arg \min } \mathcal{L}_{\textit{s}}\left(\theta_{e}, \theta_\textit{s}\right)-\lambda_1 \mathcal{L}_{\textit{c}}(\theta_{e}, \hat{\theta}_\textit{c})-\lambda_2 \mathcal{L}_{\textit{d}}(\theta_{e}, \hat{\theta}_\textit{d}),\\
(\hat{\theta}_\textit{c}) &=\underset{\theta_\textit{c}}{\arg \min } \mathcal{L}_{\textit{c}}(\hat{\theta}_{e}, \theta_\textit{c}), \\
(\hat{\theta}_\textit{d}) &=\underset{\theta_\textit{d}}{\arg \min } \mathcal{L}_{\textit{d}}(\hat{\theta}_{e}, \theta_\textit{d}).
\end{aligned}
\end{equation*}

\begin{table*}[]
\caption{Details and results of our submitted UR-AIR systems for DF and LA tasks. Both systems are fused by several sub-systems. 
} 
\setlength{\tabcolsep}{6pt}
\renewcommand{\arraystretch}{1.1}
  \centering
\begin{tabular}{P{15pt}P{20pt}P{40pt}P{55pt}P{50pt}P{50pt}P{28pt}P{28pt}P{30pt}P{40pt}}
\hline\hline
\multirow{2}{*}{\textbf{Task}} &\multirow{2}{*}{\textbf{Index}} &\multirow{2}{*}{\textbf{Feature}} &\multirow{2}{*}{\textbf{Backbone}}&\multirow{2}{*}{\textbf{Loss}} &  \multirow{2}{*}{\textbf{Strategy}} &
\multicolumn{2}{c}{\textbf{Augmentation}} &
\multicolumn{2}{c}{\textbf{Results}}\\ 
\cline{7-9} \cline{9-10}
&&&&&&Codec&Device&EER&min-tDCF\\
\hline
\multirow{5}{*}{\textbf{DF}}&1&\multirow{5}{*}{LFCC}&ECAPA-3&\multirow{5}{*}{OC-Softmax} & AUG &\ding{51}&\ding{55} &\multirow{5}{*}{\textbf{20.33}}&\multirow{5}{*}{N/A}\\
&2& &ECAPA-3 & &AUG  &\ding{51}&\ding{51}& &\\
&3& &ECAPA-3 & & ADV-AUG  &\ding{51}&\ding{51}& &\\
&4&  &ECAPA-3 && ADV-AUG  &\ding{51}&\ding{55}& & \\
&5&   & ResNet&& ADV-AUG  &\ding{51}&\ding{51}& & \\
\hline

\multirow{7}{*}{\textbf{LA}}&1&\multirow{7}{*}{LFCC}&ECAPA-2& OC-Softmax  & AUG  &\ding{51}&\ding{55}&\multirow{7}{*}{\textbf{5.46}}&\multirow{7}{*}{\textbf{0.3094}} \\
&2&&ECAPA-1& OC-Softmax&AUG  &\ding{51}&\ding{55} & &\\

&3&&ECAPA-1& P2SGrad& AUG  &\ding{51}&\ding{55} &&\\

&4&&ECAPA-3 & OC-Softmax&ADV-AUG  &\ding{51}&\ding{55}& &\\
&5&&ECAPA-3&OC-Softmax& AUG  &\ding{51}&\ding{55}& &\\
&6&&ECAPA-3&OC-Softmax& AUG  &\ding{51}&\ding{51}& & \\
&7&&ECAPA-3&OC-Softmax& ADV-AUG  &\ding{51}&\ding{51}& & \\
\hline\hline
\end{tabular}
\label{tab:results}
\end{table*}

\section{Experimental Setup}
\label{sec:exp}
\subsection{Dataset}
\textbf{Training Data.}
According to the rule of the challenge, we are only allowed to use the training and development partitions of the ASVspoof 2019 database~\cite{wang2020asvspoof} to develop our model. No external speech data is allowed, but non-speech data such as impulse responses are allowed. For both LA and DF tasks, we use 2019train as the source data and perform data augmentation with methods shown in Section~\ref{ssec:dataaug}. 

For each of the augmentation methods mentioned in Section~\ref{sssec:codec}, codec parameters such as bit rate, packet loss are randomly sampled to handle different settings. This results in 60 transmission codec options for LA and 24 compression codec options for DF. For transmission only and device-transmission paired degradation pipelines, we randomly sampled and created 21 paralleled transmission codec audio files and 6 paralleled compression codec audio files for LA and DF, respectively. For the additional device augmentation step, we sampled 12 device impulse responses including iPhone, iPad, telephone, and several different microphones.

\textbf{Evaluation Data.}
The challenge uses a new set of evaluation data for each task. For the DF task, the data was primarily generated by ASVspoof2019 LA eval data going through several unknown compression codec channels, which result in a total of 611,829 trials. As for the LA task, the TTS, VC and hybrid spoofing attacks were the same as those from the ASVspoof 2019 LA evaluation partition. Similarly, it was degraded by different unknown transmission channels, totaling 181,566 trials.

\subsection{Feature Extraction}
Following our previous works~\cite{zhang2021one, zhang2021empirical}, we extract 60-dimensional linear frequency cepstral coefficients (LFCCs) with a 20ms frame size and 10ms hop size. To form batches, we set a fixed length as exactly 750 frames. For shorter utterances, we apply repeat padding until the utterance is longer than the fixed length. Then we randomly truncate along the time axis to choose a consecutive sequence. 

\subsection{Training Details}
After feature extraction from augmented data, we train several sub-systems with Adam optimizer to update the weights in models. Each training process is run on a single NVIDIA GTX 1080 Ti GPU with a learning rate of 0.0005, over 200 epochs, and a batch size of 64. The learning rate is reduced by half every 30 epochs. We set $\lambda_1$ and $\lambda_2$ in the ADV-AUG training strategy as 0.05. Then we select the model with the lowest validation loss for evaluation. The implementation is available at \href{https://github.com/yzyouzhang/ASVspoof2021_AIR}{\texttt{github.com/yzyouzhang/ASVspoof2021\_AIR}}.

\subsection{Fused Systems}
We perform the logistic regression with Bosaris~\cite{brummer2013bosaris} toolkit to fuse several sub-systems. We choose several ECAPA-TDNN based systems for the fusion stage. Besides the primary models we described in Section~\ref{ssec:model}, we also fuse a secondary ResNet system shown in Table~\ref{tab:results}. 
In the LA task, we also include a system with P2SGrad loss function proposed in~\cite{wang2021comparative}, which achieves 4.77\% EER and min-tDCF 0.2545 in the progress phase\footnote{In the progress phase, only part of the data is used to calculate the metrics. The ground truth label is withheld in both the progress phase and evaluation the phase.}, although it is not stable in other settings. The final fused system is presented in Table~\ref{tab:results}. 

\subsection{Evaluation Metrics}


The output score of the SSD system indicates the confidence of the utterance belonging to the bona fide speech. 
Equal Error Rate (EER) is calculated by setting a threshold on the output score such that the miss probability is equal to the false alarm probability. The lower EER denotes the better discrimination ability of the SSD system. 
Tandem detection cost function (t-DCF)~\cite{kinnunen2020tandem} assesses the influence of anti-spoofing systems on the reliability of an ASV system. To compare different anti-spoofing systems, the ASV system is fixed in the challenge. The lower t-DCF indicates the better reliability of the SSD system on ASV. The t-DCF is slightly revised compared to the ASVspoof 2019 challenge. 

Based on the evaluation plan of the ASVspoof2021 challenge, the metric for the DF task is EER. 
For the LA task, the main metric is t-DCF, but we also include EER in our analysis. 

\section{Results and Analysis}
\label{sec:res}





\begin{table*}[!htbp]
\setlength{\tabcolsep}{6pt}
\caption{Comparison of EER (\%) performance pooled over attacks on DF.}
\renewcommand{\arraystretch}{1.1}
\centering
\begin{tabular}{c|ccccccccc|c}
\hline\hline
System & DF-C1  &  DF-C2  &  DF-C3  &  DF-C4  &  DF-C5  &  DF-C6  &  DF-C7   & DF-C8  &  DF-C9  &  Pooled \\
\hline
CQCC-GMM	 & 19.48  &  48.86  &  20.37  &  \textbf{19.55}  &  \textbf{20.27}  &  17.92  &  \textbf{14.42}  &  49.41  &  17.39  &  25.56    \\
LFCC-GMM	 & \textbf{17.39}  &  39.20  &  \textbf{17.97}  &  20.95  &  21.43  &  22.16  &  14.83  &  39.03  &  26.65  &  25.25    \\
LFCC-LCNN	 & 23.19  &  34.21  &  23.88  &  25.22  &  23.85  &  19.06  &  17.10  &  28.35  &  18.54  &  23.48    \\
RawNet2	 & 26.98  &  \textbf{27.63}  &  27.49  &  26.72  &  27.23  &  18.80  &  18.67  &  \textbf{18.74}  &  19.10  &  22.38    \\
\hline
\textbf{Ours}	 & 22.04  &  29.56  &  24.76  &  22.86  &  23.04  &  \textbf{16.04}  &  15.74  &  19.54  &  \textbf{16.17}  &  \textbf{20.33}    \\
\hline\hline
\end{tabular}
\label{tab:pool_DF}
\end{table*}

\begin{table*}[!htbp]
\setlength{\tabcolsep}{6pt}
\caption{Comparison of EER (\%) performance pooled over attacks on LA.}
\renewcommand{\arraystretch}{1.1}
\centering
\begin{tabular}{c|ccccccc|c}
\hline\hline
System & LA-C1  &  LA-C2  &  LA-C3  &  LA-C4  &  LA-C5  &  LA-C6  &  LA-C7 &  Pooled \\
\hline
CQCC-GMM	 & 10.57 &  14.76 & 20.58 & 11.61 & 13.58 & 14.01 & 11.21 & 15.62     \\
LFCC-GMM	 & 12.72  &  21.21  &  35.55  &  15.28  &  18.76  &  18.46  &  12.73  &  19.30    \\
LFCC-LCNN	 & 6.71  &  8.89  &  12.02  &  6.34  &  9.25  &  11.00  &  6.66  &  9.26    \\
RawNet2	 & 5.84  &  6.59  &  16.72  &  6.41  &  6.33  &  10.66  &  7.95  &  9.50    \\
\hline
\textbf{Ours}	 & \textbf{4.73}  &  \textbf{4.47}  &  \textbf{6.20}  &  \textbf{5.06}  &  \textbf{4.71}  &  \textbf{5.18}  &  \textbf{5.82}  &  \textbf{5.46}    \\
\hline\hline
\end{tabular}
\label{tab:pool_LA}
\end{table*}

\begin{table*}[!htbp]
\setlength{\tabcolsep}{6pt}
\caption{Comparison of min-tDCF performance pooled over attacks on LA.}
\renewcommand{\arraystretch}{1.1}
\centering
\begin{tabular}{c|ccccccc|c}
\hline\hline
System & LA-C1  &  LA-C2  &  LA-C3  &  LA-C4  &  LA-C5  &  LA-C6  &  LA-C7 &  Pooled \\
\hline
CQCC-GMM	 & 0.2858  & 0.5116 &  0.6183 &  0.3715 &  0.4703 &  0.4791 &  0.2935 &  0.4974    \\
LFCC-GMM	 & 0.3615 &  0.6334  & 0.8307 &  0.4409  & 0.5819 &  0.6061  & 0.3661  & 0.5758    \\
LFCC-LCNN	 & 0.2186  & 0.3354 &  0.4286 &  0.2218 &  0.3505 &  0.4217  & \textbf{0.2166}  & 0.3445     \\
RawNet2	 &  0.2073  & 0.3584  & 0.6367  & 0.2387  & 0.3489  & 0.4959  & 0.2908  & 0.4257     \\
\hline
\textbf{Ours}	 & \textbf{0.1902}  & \textbf{0.2911} &  \textbf{0.3720}  & \textbf{0.2104} & \textbf{0.2996} & \textbf{0.3618}  & 0.2186  & \textbf{0.3094}     \\
\hline\hline
\end{tabular}
\label{tab:pool_LA_tdcf}
\end{table*}

\subsection{Performance Analysis of Our Submitted System}

\begin{figure}[h]
  \centering
  \includegraphics[width=\linewidth]{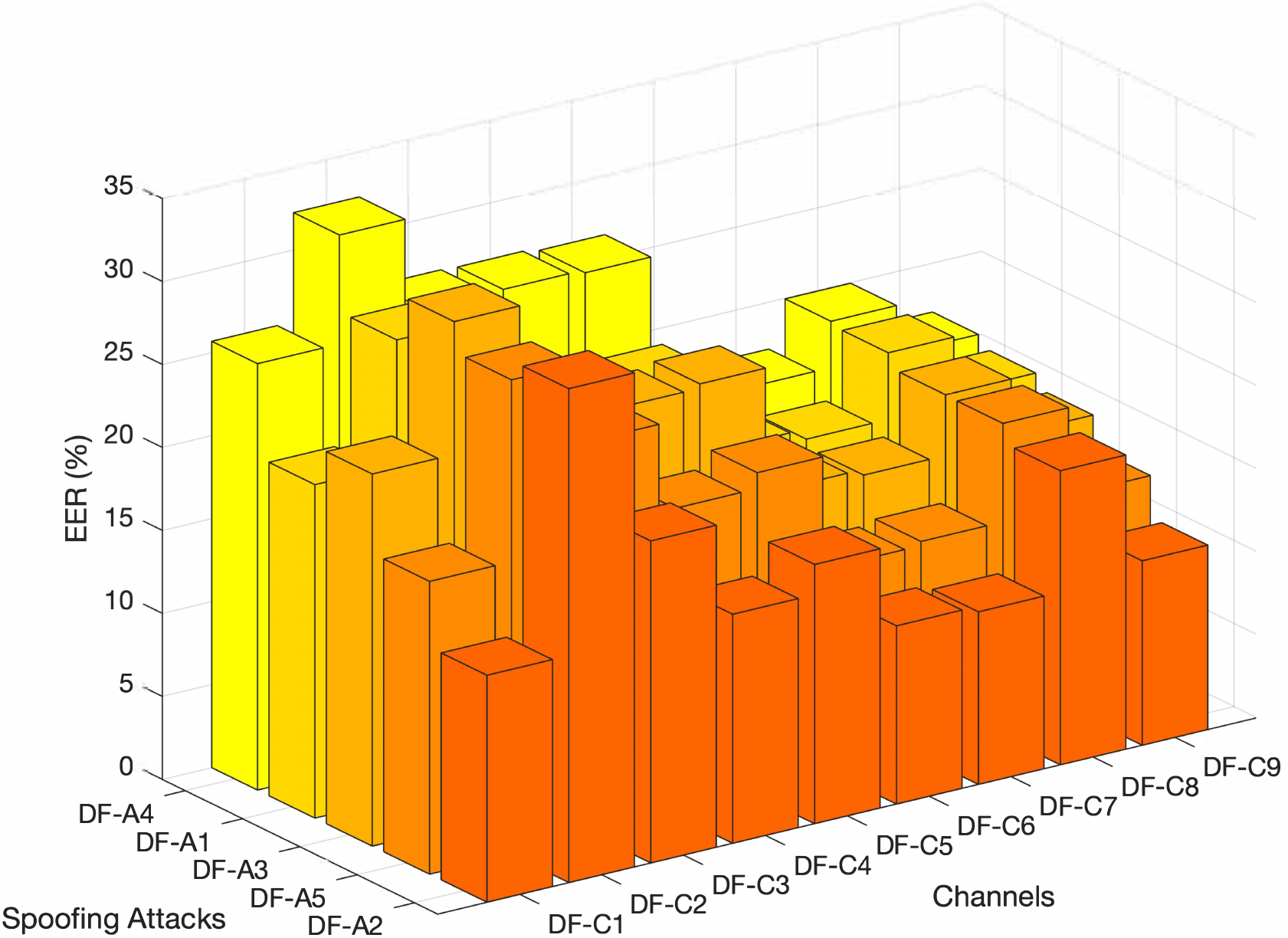}
  \caption{Detailed performance of DF submission on different channels and attacks. The spoofing attacks are traditional vocoder (A1), waveform concatenation (A2), autoregressive neural vocoder (A3), non-autoregressive neural vocoder (A4), unknown (A5). C1-C9 represents different compression codecs.
}
  \label{fig:bar_DF}
\end{figure}

\begin{figure}[h]
  \centering
  \includegraphics[width=0.95\linewidth]{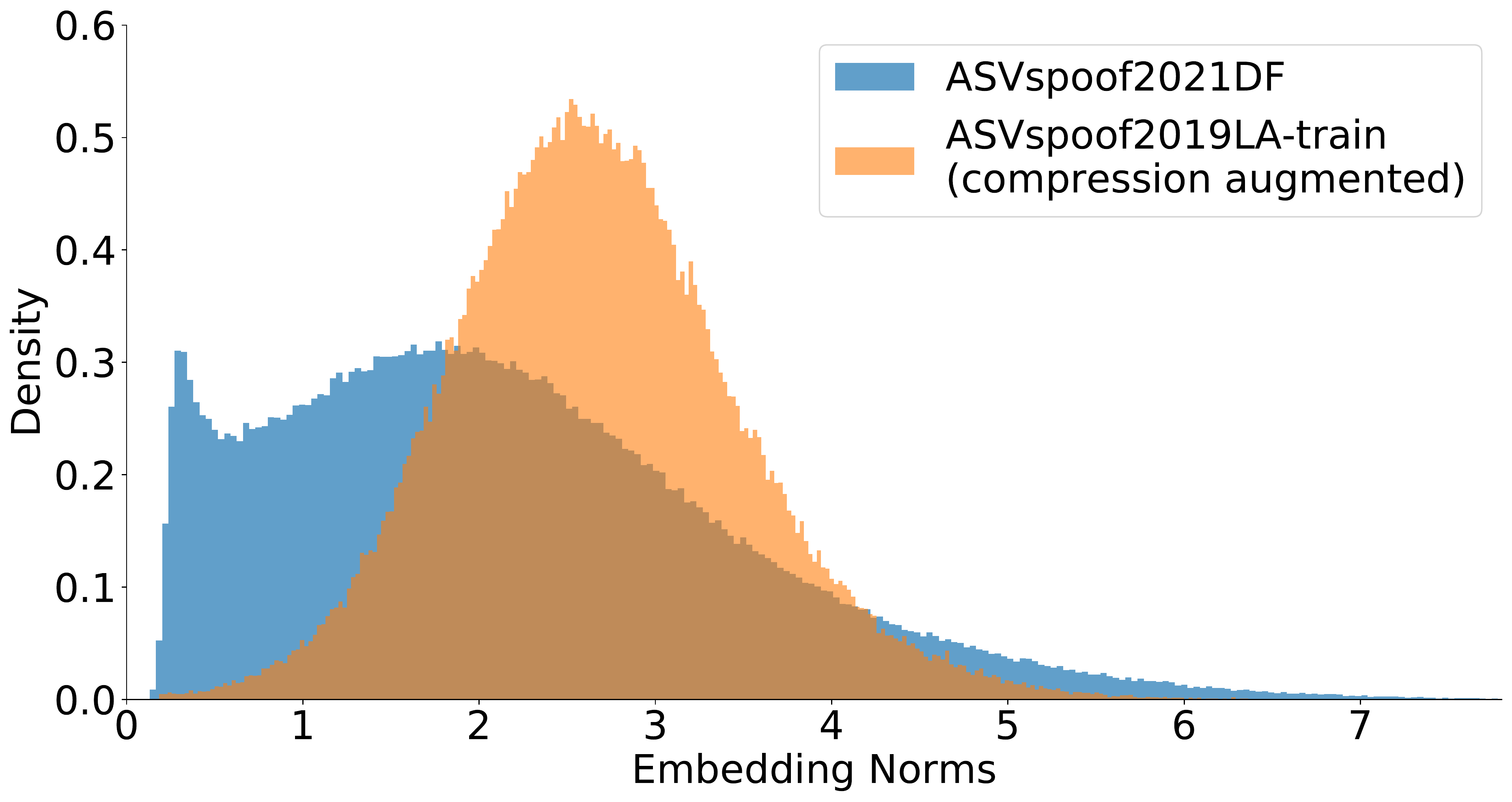}
  \caption{Density histogram of the calculated norms of the feature embeddings of one single system across the training and evaluation data for DF. 
}
  \label{fig:norm}
\end{figure}

The overall performance in the evaluation phase is shown in Table~\ref{tab:results}. The DF detailed results are reported in Figure~\ref{fig:bar_DF} with different channels and attacks. Among different channels, we observe that DF-C2 (low-quality MP3) is the most difficult for our system. While from the perspective of spoofing attacks, non-autoregressive neural vocoder is the hardest one to detect. 

In Softmax loss-based systems for verification tasks, it has been found that the learned embeddings tend to correlate their norms with the quality of input features~\cite{Ranjan17L2Constrained}: the lower the input quality is, the smaller its embedding norm is. In this paper, we also observe the same problem in the DF task. Figure~\ref{fig:norm} shows an example norm histogram of the feature vectors of the training set and evaluation set of speech utterances with different compression codecs. We can observe a significant distribution mismatch between the two datasets, which can possibly explain this poor performance in the DF task. 

\begin{figure}[h]
  \centering
  \includegraphics[width=\linewidth]{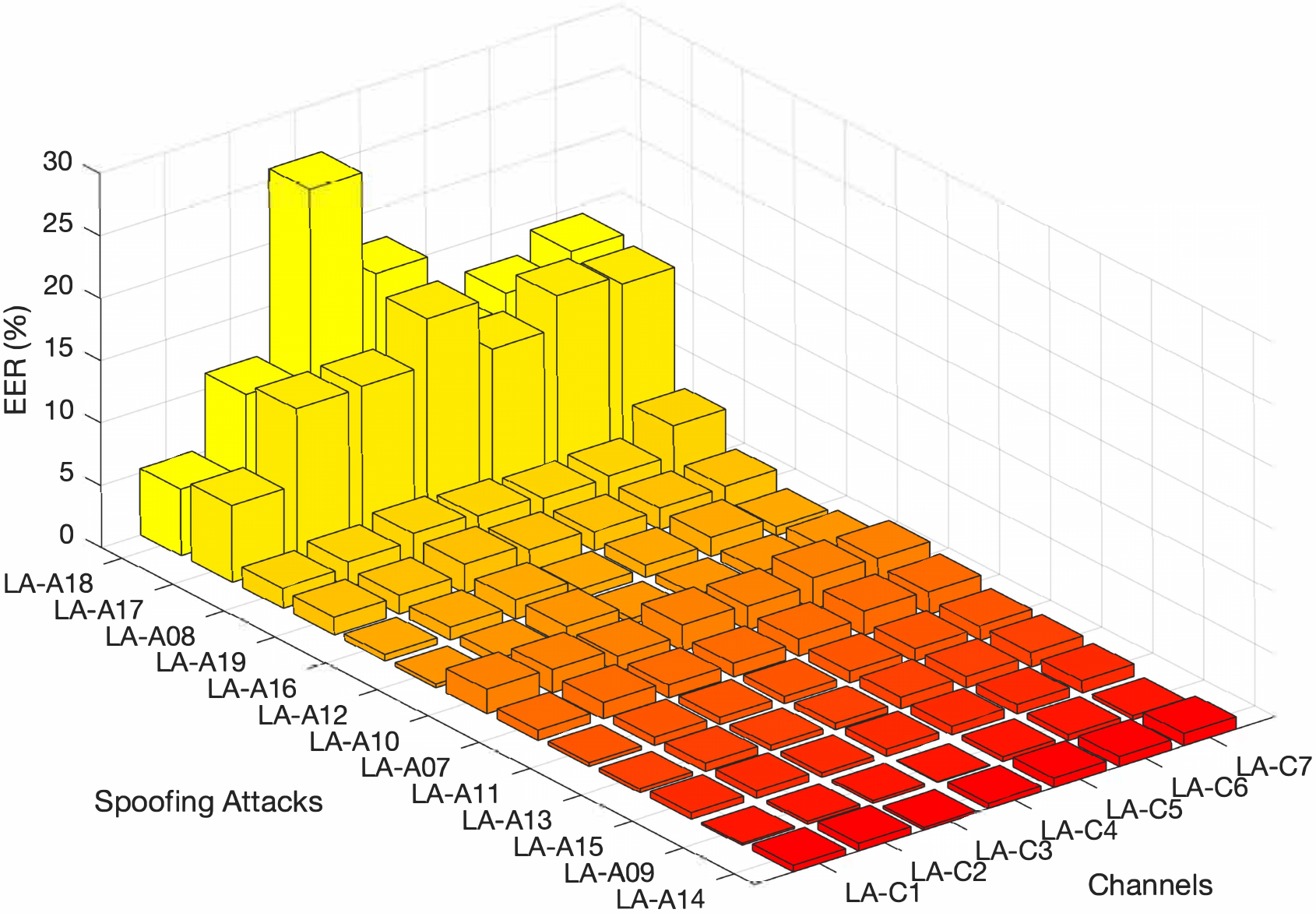}
  \caption{Detailed performance of LA submission on different channels and attacks. 
  A07-A19 represents different spoofing attacks in the evaluation data, as defined in the ASVspoof2019 database. C1-C7 represents different codec effects.
}
  \label{fig:bar}
\end{figure}

As for LA, the detailed results are reported in Figure~\ref{fig:bar} based on different waveform generation methods: vocoder (A07, A09, A14, A18), neural waveform (A08, A10, A12, A15), waveform filtering (A13, A17), spectral filtering  (A19) and waveform concatenation (A13, A16)~\cite{zeinali2019detecting}. In accordance with the analysis of the generalization performance in the ASVspoof2019 database~\cite{todisco2019asvspoof, wang2020asvspoof} where A17 was considered as the worst-case scenario, 
our system still could not improve upon this spoofing algorithm. 
Besides, we also find it difficult for our system to detect A18, 
especially transmitted through LA-C3 (public switched telephone network (PSTN)). It could be attributed to that PSTN is not included in our transmission codec augmentation. We believe that our augmentation with landline codec simulates only part of the PSTN process. 
However, even for LA-C1, i.e., no codec, our system shows a higher EER compared to our previous work~\cite{zhang2021one}. This puzzles us and further investigations are needed after we have access to the complete information of the test data.

\subsection{Comparison with Baseline Systems}
We compare our submitted system with the four baseline systems  LFCC-GMM~\cite{sahidullah2015comparison}, CQCC-GMM~\cite{todisco2016new}, LFCC-LCNN~\cite{wang2021comparative}, RawNet2~\cite{tak2021end}, provided by the organizers.
Table~\ref{tab:pool_DF} and Table~\ref{tab:pool_LA}  present the EER performance of all baseline systems and our proposed system for DF and LA respectively. For the DF task, we achieve better overall performance. In individual DF channels, it achieves comparable results on DF-C2, DF-C7, and DF-C8, and marginally improves baseline systems on DF-C6 and DF-C9. For the LA task, our proposed system performs better than all of the baseline systems and is more robust, resulting in evenly distributed low EERs on all spoofing conditions. We also present min-tDCF performance for LA in Table~\ref{tab:pool_LA_tdcf}.
\subsection{Model Comparison}

Table~\ref{tab:mwsystems} compares different backbone models in the progress phase. By comparing the theoretical amount of multiply-accumulate (MAC) operations among every sub-system and the number of parameters, we can easily find that LCNN has both the least MACs and parameters while ECAPA-4 and ECAPA-6 have the largest. Although more GPU capacity and time are needed for training ECAPA models, they demonstrated lower EERs as well as min-tDCF compared to either LCNN or Res2Net. By comparing a small-scale ECAPA-TDNN (C=512) with a large-scale version (C=1024), we observe that the smaller versions outperform the larger ones which have over twice of parameters and MACs, indicating that large models have already over-fitted on the training data. By comparing systems ECAPA-1 and ECAPA-2 with ECAPA-3, we can see that the improvement of context vector before pooling is as significant as MFA in anti-spoofing task in terms of min-tDCF.
\begin{table}[ht!]
\setlength{\tabcolsep}{4.5pt}
\renewcommand{\arraystretch}{1.1}
  \centering
  \caption{Comparison of different backbone models on the ASVspoof 2021 LA task in the progress phase.}
  \begin{tabular}{P{38pt}|P{40pt} P{32pt} | P{25pt} P{40pt}}
    \hline\hline
    \textbf{System}    & \textbf{Parameters}& \textbf{MACs}& \textbf{EER} & \textbf{min-tDCF}\\
    \hline
    LCNN& 0.52 M & 0.63 G &  7.28 & 0.2738\\
    Res2Net& 0.89 M & 1.47 G & 7.28 & 0.2801\\
     \hline
    ECAPA-1& 6.34 M & 3.87 G  & 5.31  & 0.2732\\
    ECAPA-2& 5.94 M & 3.57 G & 5.14  &0.2759 \\
    ECAPA-3& 6.34 M & 3.87 G & 5.13 & 0.2626\\
    ECAPA-4& 14.75 M & 9.89 G  & 5.77&0.2725 \\
    ECAPA-5& 14.36 M & 9.59 G & 6.39   & 0.2818\\
    ECAPA-6& 14.75 M & 9.89 G & 6.93& 0.2942\\
    \hline\hline
  \end{tabular}
  \label{tab:mwsystems}
\end{table}

\subsection{Effect of Model Fusion}
\begin{figure}[h]
  \centering
  \includegraphics[width=0.66\linewidth]{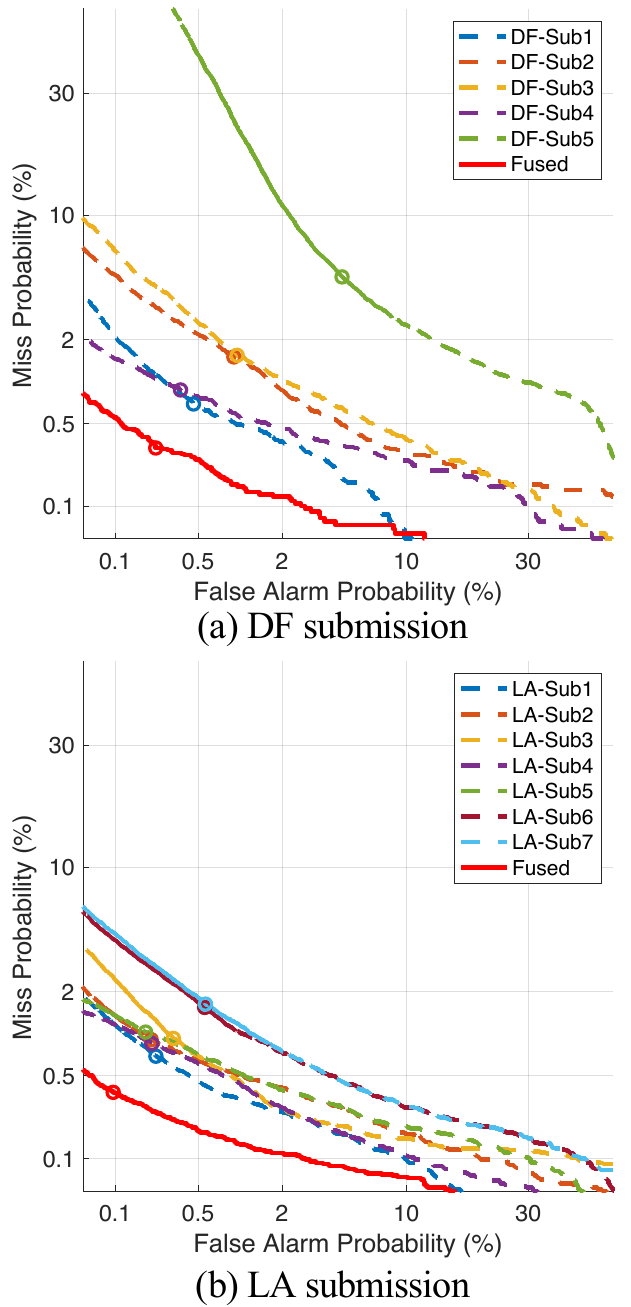}
  \caption{Performance of model fusion of our submissions on the augmented development set. Dashed curves represent the DET curve of individual systems and the red solid curve represents that of our fused system. The circles on the DET curves represent the operating points of the min-tDCF. The indices of the sub-systems are consistent with those in Table~\ref{tab:results}.}
  \label{fig:fusion}
\end{figure}

In Figure~\ref{fig:fusion}, we plot the detection error tradeoff (DET) curve to show the miss probability and false alarm probability trade-off by varying the threshold for the output synthetic speech classification score. 
We compute a DET curve for each individual system on the codec-augmented ASVspoof2019LA-dev dataset during the progress phase.
From Figure~\ref{fig:fusion}, it is obvious that the fused model strongly outperforms individual ones in both DF and LA tasks. Hence, we submit our fused system during the evaluation phase.
Note that the conclusion that model fusion is better than individuals is drawn from the performance on the augmented development set. We believe further investigations are needed on the ASVspoof 2021 evaluation set during the post-evaluation phase. 

\section{Conclusions}
\label{sec:sum}
In this paper, we introduced UR-AIR system for the channel-robust anti-spoofing task. Specifically, to improve the robustness over channel variability, we use an acoustic simulator to apply transmission codec, compression codec, and convolutional device impulse responses to the training data. To learn a reliable embedding from the augmented dataset, we use several variants of ECAPA-TDNN as our backbone neural network architecture. We also incorporate one-class learning with channel-robust training strategies to further learn a channel-invariant speech representation. Our submission achieved considerable improvement over baseline systems in both DF and LA tasks of the ASVspoof 2021 challenge.

\section{Acknowledgments}
This work was supported by National Science Foundation grant No. 1741472 and funding from Voice Biometrics Group. You Zhang would like to thank the synergistic activities provided by the NRT program on AR/VR funded by NSF grant DGE-1922591. The authors would also like to thank the organizers for providing additional results breakdown for our submissions.

\bibliographystyle{IEEEbib}
\bibliography{ASVspoof2021_BibEntries}

%

\end{document}